\documentclass[twocolumn]{article}

\usepackage{amsmath,amssymb,amsfonts}
\usepackage{amsthm}
\usepackage{hyperref}
\usepackage{algorithm}
\usepackage{algpseudocode}

\DeclareFontFamily{U}{mathx}{}
\DeclareFontShape{U}{mathx}{m}{n}{<-> mathx10}{}
\DeclareSymbolFont{mathx}{U}{mathx}{m}{n}
\DeclareMathAccent{\widehat}{0}{mathx}{"70}
\DeclareMathAccent{\widecheck}{0}{mathx}{"71}
\newtheorem*{lem*}{Lemma}
\usepackage{graphicx, color, caption, subcaption}
\graphicspath{{fig/}}
\usepackage{mathtools}

\DeclarePairedDelimiterX{\infdivx}[2]{(}{)}{%
	#1\;\delimsize\|\;#2%
}

\usepackage{pgf}
\usepackage{bm}

\newtheorem{lem}{Lemma}

\newcommand{\guillemets}[1]{``#1''}
\newcommand{\set}[1]{\left\{#1\right\}}

\newcommand{\myvector}[1]{\bm{#1}}

\newcommand{\mymatrix}[1]{\bm{#1}}
\newcommand{\entry}[1]{\left[{#1}\right]}
\newcommand{\hermitian}{H}

\newcommand{\R}{\mathbb{R}} % Ensemble R
\newcommand{\C}{\mathbb{C}} % Ensemble C
\newcommand{\E}{\mathrm{E}} % Espérance

\newcommand{\eqdef}{\triangleq}

\newcommand{\norm}[1]{\left\lVert#1\right\rVert}
\newcommand{\abs}[1]{\left|#1\right|}

\newcommand{\ie}{{i.e.},}
\newcommand{\eg}{{e.g.},}

\newcommand{\toas}{\xrightarrow{\mathrm{a.s.}}}

\DeclareMathOperator{\Real}{\mathrm{Re}}
\DeclareMathOperator{\diag}{\mathrm{diag}}
\DeclareMathOperator*{\vect}{\mathrm{vec}}
\DeclareMathOperator*{\trace}{tr}

\DeclareUnicodeCharacter{2212}{-}

\begin{document}

\title{Moment-Matching Array Processing Technique for diffuse source estimation} %% Article title

\author{Colin Cros\textsuperscript{1}, Laurent Ferro-Famil\textsuperscript{1,2}\\
	\textsuperscript{1} ISAE Supaero, University of Toulouse, France \\
	\textsuperscript{2} CESBIO, University of Toulouse, France \\
	\texttt{firstname.lastname@isae-supaero.fr}
}

\maketitle

%% Abstract
\begin{abstract}
%% Text of abstract
	Direction of Arrival (DOA) estimation is a fundamental problem in signal processing. Diffuse sources, whose power density cannot be represented with a single angular coordinate, are usually characterized based on prior assumptions, which associate the source angular density with a specific set of functions. However, these assumptions can lead to significant estimation biases when they are incorrect. This paper introduces the Moment-Matching Estimation Technique (MoMET), a low-complexity method for estimating the mean DOA, spread, and power of a narrow diffuse source without requiring prior knowledge on the source distribution. The unknown source density is characterized by its mean DOA and its first central moments, which are estimated through covariance matching techniques which fit the empirical covariance of the measurements to that modeled from the moments. The MoMET parameterization is robust to incorrect model assumptions, and numerically efficient. The asymptotic bias and covariance of the new estimator are derived and its performance is demonstrated through simulations.
\end{abstract}

\section{Introduction}

Direction-of-arrival (DOA) estimation is central to many array processing applications, including acoustics, radar, wireless communications, underwater sonar, and synthetic aperture radar (SAR) imaging. Classical high-resolution algorithms, such as MUSIC \cite{schmidt1986multiple} or ESPRIT \cite{roy1986esprit}, characterize a small number of point-like sources using line spectrum estimation \cite{stoica2005spectral}.
When a source consists of a set of numerous radiating elements densely distributed in space \cite{aguilera2013wavelet, yue2017three} this assumption does not hold, and these techniques lead to erroneous results \cite{jantti1992influence}. Several modifications to the MUSIC and ESPRIT algorithms have been proposed to solve the DOA estimation problem in the case of diffuse sources \cite{valaee1995parametric, meng1996estimation, wu1994estimation, shahbazpanahi2001distributed}. However, these subspace methods may face severe limitations in the presence of full-rank source signals. Weighted pseudo-subspace fitting (WPSF) \cite{bengtsson2001generalization} has been proposed to overcome this difficulty, but it comes at the cost of higher computational complexity.

Another limitation of these subspace methods is their need for a source distribution model, which is typically assumed to belong to a specific class of functions. When the shape of a diffuse source density remains unknown, setting an \textit{a priori} candidate might induce model misspecifications, and lead to estimation biases, as this will be demonstrated. Model misspecifications in DOA estimation have been studied from different perspectives, such as, for instance, the incorrect modelling of array manifolds \cite{swindlehurst1992performance, ferreol2010statistical}, or misspecified noise statics \cite{tsakalides1996robust, tsung2001subspace, tian2019robust, fortunaty2016misspecified}. % However, to the best of the authors' knowledge, no specific algorithm has been proposed to identify sources whose angular distribution is unknown.
A second-order approximation of the covariance matrix has been proposed to create a low-complexity estimation algorithm \cite{bengtsson2000lowcomplexity}, though it still uses a model to estimate the spread. The use of higher orders has also been explored: in \cite{shahbazpanahi2004covariance}, an iterative algorithm is proposed to identify several diffuse sources based on a Taylor expansion of their distributions and estimation of their non-central moments. However the estimation of the source spread still relies on the use of an \textit{a priori} shape.
Finally, methods that do not rely on assumptions about the source distribution have been proposed for monopulse radar \cite{milstein1981maximum, monakov2012maximum, nickel2013estimation}, although those methods do use prior information about the source's position.

This paper introduces a Moment-Matching Estimation Technique (MoMET) to estimate three relevant characteristics of a narrow diffuse source: its mean DOA, spread, and power. The MoMET algorithm does not require any prior information about the source distribution. Instead, the diffuse source density is parameterized by its moments.

The main contributions are the following:
\begin{enumerate}
	\item A model for the covariance matrix of a diffuse source based on the central moments of the source's distribution is introduced.  This model does not require knowledge of the source power density's shape, and is linear in all parameters except the mean DOA.
	\item A low-complexity estimation algorithm, called MoMET, is derived by fitting the proposed model to the sample covariance matrix. This algorithm can be applied to both uniform linear arrays (ULAs) and non-ULAs.
	\item The performance of MoMET is derived in terms of bias and variance, and is validated numerically. Specifically, increasing the number of estimated moments is proved to improve the estimation performance.
\end{enumerate}
This work follows on from a preliminary conference communication by the authors \cite{cros2025moment}. However, the results presented here are more detailed, self-contained, and contain new materials; for example; no performance studies or problem analyses were conducted in \cite{cros2025moment}. While the development of MoMET was motivated by forest SAR tomography applications \cite{aghababaei2020forest} in which forest density of reflectivity does not follows any simple model, MoMET can be applied to any DOA estimation problem.

The rest of the paper is organized as follows. Section~\ref{sec: Problem formulation} introduces and discusses the estimation problem, then Section~\ref{sec: Moment-based estimation} presents the new estimation method. Section~\ref{sec: Performance analysis} examines the theoretical performance, and Section~\ref{sec: Numerical example} assesses this performance with simulations.

\bigbreak
\textbf{Notation.}
In the sequel, lowercase boldface letters denote vectors \eg{} $\myvector{x} \in \C^M$, and uppercase boldface variables represent matrices \eg{} $\mymatrix{A} \in \C^{M\times N}$. The trace, inverse, transpose, Hermitian, and the  $(k,l)$ entry of a matrix $\mymatrix{A}$ are denoted by $\trace \mymatrix{A}$, $\mymatrix{A}^{-1}$, $\mymatrix{A}^{\intercal}$, $\mymatrix{A}^{\hermitian}$, and $\entry{\mymatrix{A}}_{k,l}$, respectively. $\mymatrix{I}_M$ is the ($M\times M$) identity matrix. The Hadamard (element-wise) product and the Kronecker product of two matrices are denoted by $\mymatrix{A} \odot \mymatrix{B}$ and $\mymatrix{A} \otimes \mymatrix{B}$, respectively. For a matrix $\mymatrix{A} \in \C^{M \times N}$, $\vect{} \mymatrix{A} \in \C^{MN}$ is the vector obtained by stacking the columns of $\mymatrix{A}$. For a vector $\myvector{a} \in \C^M$, $\diag \mymatrix{a}\in\C^{M\times M}$ is the diagonal matrix induced by $\myvector{a}$. The Loewner partial order on positive semi-definite matrices is denoted by $\preceq$.
The notation $\E[\cdot]$ denotes the expected value of a random variable and $\norm{\cdot}$ the Euclidean norm.

\section{Problem formulation}\label{sec: Problem formulation}

Consider a $2$D space containing a linear array consisting of $M$ sensors, measuring the signal emitted by a distributed source, covering a narrow spectral band.
The signals received by the $M$ elements of the array are gathered into a vector, denoted by $\myvector{x}(t) \in \C^M$, modeled as:
\begin{equation}\label{eq: Received signal}
	\myvector{x}(t) =  \int \myvector{a}(\omega)s(\omega, t) \, d\omega + \myvector{\epsilon}(t).
\end{equation}
In \eqref{eq: Received signal}, $t$ stands for the acquisition index, $s(\omega, t)$ denotes the source signal density at angular position $\omega$, $\myvector{\epsilon}(t)$ stands for the acquisition noise, and $\myvector{a}(\omega)$ represents the steering vector, which models the relationship between the DOA and the phase shifts measured by the array elements,
\begin{equation}
	\myvector{a}(\omega) \eqdef \begin{pmatrix}
		1 & e^{j u_2 \omega} & \dots & e^{j u_{M} \omega}
	\end{pmatrix}^{\intercal},
\end{equation}
where $u_k$ denotes the spacing between the first and the $k$-th sensor, normalized by half the carrier wavelength.

The expression given in  \eqref{eq: Received signal} is general and can model signals acquired using any array configuration. In the case of an ULA, with $u_k > u_{k-1}$, the array resolution, $\delta_\omega$, and ambiguous domains, $\Delta  \omega_a$, are given by:
\begin{align*}
\delta_\omega &\eqdef\frac{2 \pi}{M du},&  \Delta  \omega_a &\eqdef \min \left(2 \pi,\frac{2 \pi}{du} \right),
\end{align*}
respectively, with $du \eqdef \abs{u_k - u_{k-1}}$. It is assumed in the following that the array is operated in a conventional way, \ie{} the inter-element spacing and an appropriate demodulation step, required in case of squinted acquisition geometry, guarantee the measured source density is contained within the array's unambiguous angular domain, \ie{} $s(\omega,t)=0$, $\forall \abs{\omega}>\frac{\Delta \omega_a}{2}$. This hypothesis also applies to non-ULA configurations, using the appropriate resolution and ambiguity definitions provided, \eg{} in \cite{babu2010spectral}. For the sake of clarity and without loss of generality, it is assumed in the following that $du \approx 1$ (and $du = 1$ when ULAs are considered). %[Babu & STOICA]

The noise $\myvector{\epsilon}$ is assumed to follow a zero-mean complex circular normal distribution that is spatially and temporally white, thus $\forall (t,t')$:
\begin{align}
	\E\left[\myvector{\epsilon}(t) \myvector{\epsilon}(t')^{\intercal}\right] &= \mymatrix{0}, &
	\E\left[\myvector{\epsilon}(t) \myvector{\epsilon}(t')^{\hermitian}\right] &= \sigma_{\epsilon}^2 \delta(t-t')\mymatrix{I}_M,
\end{align}
where $\delta$ denotes the Dirac delta function. The source signal is assumed to be a zero-mean, stationary random process following a normal distribution. Its autocorrelation kernel is characterized by its power density $f$ such that $\forall (\omega, \omega', t, t')$:
\begin{subequations}
\begin{align}
	\E\left[s(\omega,t) s(\omega', t'))\right] &= 0, \\
	\E\left[s(\omega,t) s^*(\omega', t'))\right] &= \delta(t-t')\delta(\omega-\omega') f(\omega).
\end{align}
\end{subequations}

Therefore, the measurements $\myvector{x}(t)$ follow an unconditional model \cite{stoica1990performance}: they are i.i.d., zero-mean. Their covariance matrix is:
\begin{equation}\label{eq: Covariance matrix}
	\mymatrix{R} \eqdef \E\left[\myvector{x}(t) \myvector{x}(t)^{\hermitian}\right] 
	= \int \myvector{a}(\omega) \myvector{a}(\omega)^{\hermitian} f(\omega) \, d\omega + \sigma_\epsilon^2\mymatrix{I}_M.
\end{equation}

The objective of this study is to estimate, without any prior information on the source power density, three characteristics of the source: $(i)$ its power $P$, $(ii)$ its mean value $\omega_0$, and $(iii)$ its spread $\sigma_\omega$. They correspond to the first three moments of $f$:
\begin{subequations}\label{eq: Source characteristics}
\begin{align}
	P &\eqdef \int f(\omega) \, d\omega, \\
	\omega_0 &\eqdef \frac{1}{P}\int \omega f(\omega) \, d\omega, \\
	\sigma_\omega^2 &\eqdef \frac{1}{P} \int (\omega-\omega_0)^2 f(\omega) \, d\omega.
\end{align}
\end{subequations}
The main difficulty is that no model for the power density $f$ is available.
%Furthermore, the study is conducted in the case where the source's spread is smaller than the Fourier resolution:
%\begin{equation}\label{eq: Assumption resolution}
%	\sigma_\omega < \frac{2 \pi}{M}.
%\end{equation}

\section{Moment-Matching Estimation Technique}\label{sec: Moment-based estimation}

In order to retrieve the characteristics of the power density without any prior on the shape of the density, this section introduces a new model based on a parameterization of the power density using its central moments. The next subsection provides the motivations behind the MoMET.

\subsection{Motivation}

Estimating a power density, $f$, known to belong to a certain class of functions parameterized by some vector $\myvector{\psi}$, is a classical, well-studied unconditional DOA estimation problem, see \eg{} \cite{stoica1990performance}, whose ML solution can be derived \cite{trump1996estimation}. However, in the case of an unknown density, a more generic approach is required.

\begin{figure*}
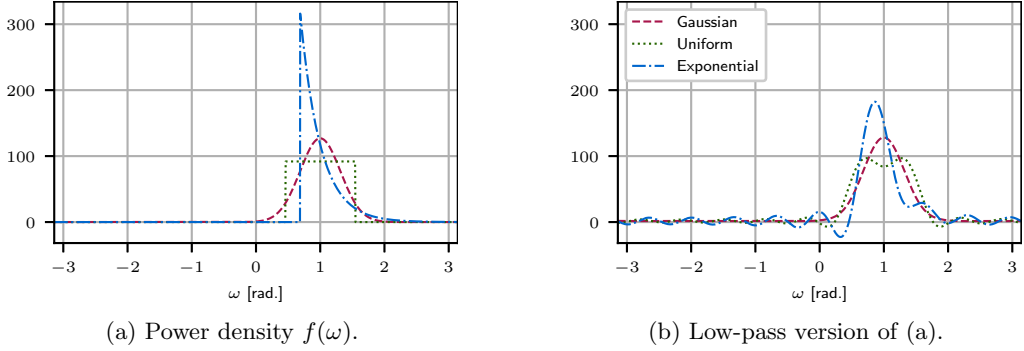

	\centering
	\null\hfill
	\subfloat[Power density $f(\omega)$.\label{sfig: Distribution}]{\input{fig/distributions.pgf}}
	\hfill
	\subfloat[Low-pass version of (a).\label{sfig: Low-pass distribution}]{\input{fig/distributions_passe_bas.pgf}}
	\hfill\null
	\caption{Comparison of three power densities and their low-pass versions computed from the entries of $\mymatrix{R}$. The parameters used are $\omega_0 = 1$~rad, $\sigma_\omega = 2\pi\times0.05$~rad, $P = 100$~W, and $\sigma_\epsilon^2 = 10$~W. The array is a ULA with $M = 10$ sensors.} 
	\label{fig: Comparison power densities}
\end{figure*}

As specified in \eqref{eq: Covariance matrix}, the entries of $\mymatrix{R}$ do not directly observe the power density $f$, but its spectrum. Indeed, the $(k,l)$ entry of $\mymatrix{R}$,
\begin{equation}
	\entry{\mymatrix{R
	}}_{k,l} = \int f(\omega) e^{j(u_k-u_l)\omega} \, d\omega + \delta_{k,l} \sigma_\epsilon^2,
\end{equation}
corresponds to the spectrum sampled at the spectral coordinate $\set{u_k-u_l}$. Therefore, for a limited number of array elements, only the low-pass version of $f$ can be reconstructed from $\mymatrix{R}$. Figure~\ref{fig: Comparison power densities} presents the low-pass reconstructions obtained with three commonly used power densities. 

Using a Dirac function as a very coarse approximation of a narrow source power density leads to a simple covariance model,
\begin{equation}\label{eq: Dirac model}
	\mymatrix{\widehat R}_\mathrm{Dirac}(\omega_0, P, \sigma_\epsilon^2) = P \myvector{a}(\omega_0) \myvector{a}(\omega_0)^{\hermitian} + \sigma_\epsilon^2 \mymatrix{I}_M.
\end{equation}
The model \eqref{eq: Dirac model} can be employed to estimate $\omega_0$, $P$ and $\sigma_\epsilon^2$ \cite{stoica1990maximum, ottersten1993exact}, but is not refined enough to measure the source spread. 

In the frequency domain, approximating the power density $f$ by a Dirac impulsion is equivalent to seek for a constant spectrum (up to a spatial shift). As $f$ is narrow, its Fourier transform is indeed \guillemets{flat} near the origin, nonetheless a constant approximation may be too crude. To allow for finer approximations and capture more complex behaviors, the proposed MoMET approximates the spectrum of $f$ by a polynomial instead of a constant. Furthermore, the Fourier transform of a distribution is a characteristic function, which encodes the distribution' moments in its Taylor expansion. For small spreads, the characteristic function is very smooth near the origin and some of its descriptors, such as the spread of the source density $\sigma_\omega$, can be estimated using a few of its first moments.
\begin{figure*}
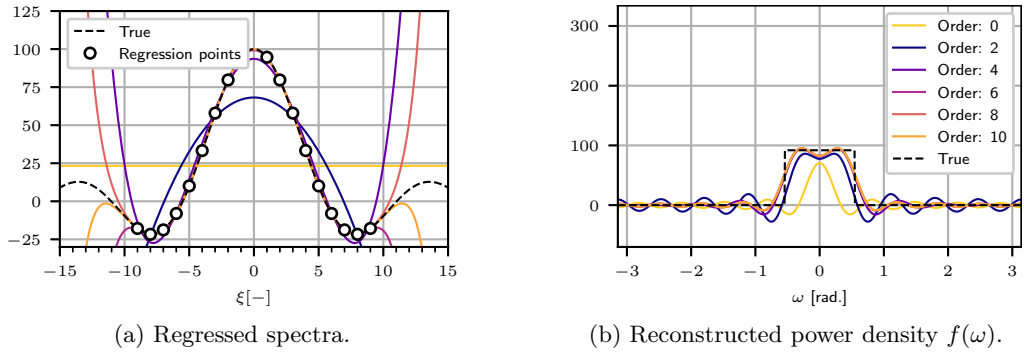

	\centering
	\null\hfill
	\subfloat[Regressed spectra.]{\input{fig/regressed_spectra.pgf}}
	\hfill
	\subfloat[Reconstructed power density $f(\omega)$.]{\input{fig/regressed_distributions.pgf}}
	\hfill\null
	\caption{Regressions of the spectrum of a uniform power density $f$ and associated reconstructions. The regression points correspond to the off-diagonal entries of $\mymatrix{R}$, $\xi_k = \sigma_\omega u_k$. The same numerical values as in Fig.~\ref{fig: Comparison power densities} have been used except $\omega_0 = 0$.}
	\label{fig: Regressions}
\end{figure*}
As an illustration, Figure~\ref{fig: Regressions} presents the reconstruction of the power density from the off-diagonal entries of a covariance matrix $\mymatrix{R}$ (since the diagonal entries do not contain any information) for different regression orders. A Dirac approximation corresponds to a $0$-th order regression. As expected, higher-order regressions produce better reconstructions and the regressed spectra tend to match the true spectrum near the origin, motivating the use of a polynomial model for the spectrum of the power density $f$.

\subsection{Covariance matrix model}

This section introduces a model, $\myvector{\widehat R}$, for the covariance matrix, $\mymatrix{R}$, based on the central moments of the power density $f$ which does not assume any prior information on the shape of the power density.

The standard distribution, $p$, is defined as a standardized version of the power density, $f(\omega)$, as:
\begin{align}\label{eq: Defintion p}
	p(x) &\eqdef \frac{\sigma_\omega}{P}f(\sigma_\omega x + \omega_0) &\Leftrightarrow\qquad f(\omega) = \frac{P}{\sigma_\omega}p\left(\frac{\omega - \omega_0}{\sigma_\omega}\right).
\end{align}
$p$ is a standard distribution in the sense that it is positive, has unit-mass, zero-mean, and unit-variance:
\begin{align}\label{eq: Constraint p}
	&\int p(x)\, dx = 1, &
	&\int x p(x)\, dx = 0, &
	&\int x^2 p(x)\, dx = 1.
\end{align}

After injecting \eqref{eq: Defintion p} into \eqref{eq: Covariance matrix}, the covariance matrix $\mymatrix{R}$ becomes:
\begin{subequations}\label{eq: Model with B}
\begin{align}
	\mymatrix{R} &= P \myvector{a}(\omega_0) \myvector{a}(\omega_0)^{\hermitian} \odot \mymatrix{B} + \sigma_\epsilon^2 \mymatrix{I}_M,\\
	\entry{\mymatrix{B}}_{k,l} &\eqdef \int p(x) e^{j (u_k-u_l) \sigma_\omega x} \, dx.\label{eq: Definition B}
\end{align}
\end{subequations}
In \eqref{eq: Model with B}, the position of the source, $\omega_0$, is decoupled from the shape of the distribution, encoded in the shaping matrix $\mymatrix{B}$. As $p$ is unknown, a model for $\mymatrix{B}$ must be set, or equivalently for the characteristic function of $p$ defined as:
\begin{equation}
	\tilde p(\xi) \eqdef \int p(x) e^{j \xi x}\, dx.
\end{equation}
Note that by definition of $\mymatrix{B}$ \eqref{eq: Definition B}, $\entry{\mymatrix{B}}_{k,l} = \tilde p((u_k-u_l)\sigma_\omega)$. It is well-known that the $d$-th derivative of the characteristic function, denoted by $\tilde p^{(d)}$, evaluated at $\xi = 0$, corresponds to the $d$-th moment of the distribution: 
\begin{align}
 	\tilde p^{(d)}(0) &= j^d \mu_d, & \mu_d &\eqdef  \int x^d p(x) \,dx.
\end{align}
Hence, for any $D$, the $D$-th order Taylor expansion at $0$ of $\tilde p$ is:
\begin{equation}
	\tilde p(\xi) = \sum_{d=0}^D \frac{j^d}{d!} \mu_d \xi^d + O(\xi^{D+1}).
\end{equation}
Since the maximum value at which $\tilde p$ is evaluated is $\xi_M = \sigma_\omega u_M$, and the exponential power series has an infinite radius of convergence, the remainder of the expansion is negligible for large $D$.

It is proposed to parameterize $\tilde p$ by retaining the first $D$ coefficients of the Taylor expansion. The choice of the expansion order $D$ is discussed in Section~\ref{sec: Performance analysis}. According to \eqref{eq: Constraint p}, the first three moments are $\mu_0 = 1$, $\mu_1 = 0$ and $\mu_2 = 1$. The moment-based model $\widehat{\tilde p}$ is then defined from a vector of new parameters,  $\myvector{\mu} = \left(\mu_3, \dots, \mu_D\right)\in \R^{D-2}$, as:
\begin{equation}\label{eq: Parameterization characteristic function}
	\widehat{\tilde p}(\xi, \myvector{\mu}) \eqdef  \sum_{d=0}^D \frac{j^d}{d!} \mu_d \xi^d = 1 - \frac{\xi^2}{2} + \sum_{d=3}^D \frac{j^d}{d!} \mu_d \xi^d.
\end{equation}
Injecting \eqref{eq: Parameterization characteristic function} into \eqref{eq: Model with B} yields the following models for $\mymatrix{R}$ and $\mymatrix{B}$:
\begin{subequations}\label{eq: Model moment brut}
\begin{align}
	\mymatrix{\widehat R}(\omega_0, P, \sigma_\omega, \myvector{\mu}, \sigma_\epsilon^2) &\eqdef P \myvector{a}(\omega_0) \myvector{a}(\omega_0)^{\hermitian} \odot \mymatrix{\widehat{B}}(\sigma_\omega, \myvector{\mu}) \notag \\
	& \qquad + \sigma_\epsilon^2 \mymatrix{I}_M,\\
	\mymatrix{\widehat{B}}(\sigma_\omega, \myvector{\mu}) &\eqdef \mymatrix{U}^{\circ 0} - \frac{\sigma_\omega^2}{2} \mymatrix{U}^{\circ 2} + \sum_{d = 3}^D  \frac{j^d \sigma_\omega^d  \mu_d}{d!}\myvector{U}^{\circ d},	
\end{align}
\end{subequations}
where $\myvector{U}^{\circ d}$ denotes the $d$-th Hadamard power of the array geometry matrix $\mymatrix{U}$ whose entries are $\left[\myvector{U}\right]_{k,l} \eqdef u_k - u_l$.

The estimation of the parameters in \eqref{eq: Model moment brut} can be simplified by introducing the vector $\myvector{\nu} = \left(\nu_2, \dots, \nu_D\right)\in \R^{D-1}$ whose entries are $\nu_d \eqdef P \mu_d \sigma_\omega^d$. As both $P$ and $\sigma_\omega$ are positive, there is a one-to-one correspondence between $(P, \sigma_\omega, \myvector{\mu})$ and $(P, \myvector{\nu})$, so estimating one or the other is equivalent. However, the parameterization $\myvector{\theta} = \left(\omega_0,\, P,\, \myvector{\nu},\, \sigma_\epsilon^2\right)$ is more convenient, as all the variables, except $\omega_0$, enter linearly in the expression of $\mymatrix{\widehat R}$. Let $\myvector{\alpha} \eqdef \left( P,\, \myvector{\nu},\, \sigma_\epsilon^2 \right)$ denote the set of linear parameters of $\myvector{\theta}$ and introduce the matrices $\set{\mymatrix{A}_k}$ associated with these parameters:
\begin{equation}\label{eq: Matrices linear parameters}
	\mymatrix{A}_k \eqdef \left\{\begin{array}{cc}
		\myvector{1}\myvector{1}^{\intercal} & \text{for the parameter $P$,}\\
		\frac{j^d}{d!}\mymatrix{U}^{\circ d} & \text{for the parameter $\nu_d$,}\\
		\mymatrix{I}_M & \text{for the parameter $\sigma_\epsilon^2$.}
	\end{array}\right.
\end{equation}
Then, grouping all the linear terms in a single matrix $\mymatrix{\widehat{B}}(\myvector{\alpha})$ leads to the model for the covariance matrix:
\begin{subequations}\label{eq: Modeled covariance}
	\begin{align}
		\mymatrix{\widehat R}(\myvector{\theta}) &\eqdef \myvector{a}(\omega_0) \mymatrix{a}(\omega_0)^{\hermitian} \odot \mymatrix{\widehat{B}}(\myvector{\alpha}), \\
		\mymatrix{\widehat{B}}(\myvector{\alpha}) &\eqdef \sum_k \alpha_k \mymatrix{A}_k=P \mymatrix{U}^{\circ 0} + \sum_{d=2}^D \nu_d \frac{j^d}{d!} \myvector{U}^{\circ d} + \sigma_\epsilon^2 \mymatrix{I}.
	\end{align}
\end{subequations}
Note that, since the Taylor expansion is truncated, the moment-based model \eqref{eq: Modeled covariance} is misspecified by design. The consequences of this misspecification are discussed in Section~\ref{sec: Performance analysis}.

\subsection{Estimation algorithm}\label{ssec: Estimation algorithm}

The MoMET algorithm aims at retrieving the parameters $\myvector{\theta}$ in \eqref{eq: Modeled covariance} using COMET estimators  \cite{ottersten1998covariance}, which are based on a generalized least squares method offering a numerically efficient alternative to the ML estimator. Introduce the cost function,
\begin{equation}\label{eq: Definition J}
	J(\myvector{\theta}) \eqdef \frac{1}{2} \norm{\mymatrix{W}^{1/2}(\mymatrix{\bar R}_N - \mymatrix{\widehat R}(\myvector{\theta})) \mymatrix{W}^{1/2}}^2 = \frac{1}{2} \norm{\mymatrix{\bar R}_N - \mymatrix{\widehat R}(\myvector{\theta})}^2_{\mymatrix{W}},
\end{equation}
where $\mymatrix{\bar R}_N \eqdef \frac{1}{N} \sum_{t=1}^N\myvector{x}(t)\myvector{x}(t)^{\hermitian}$ is the sample covariance matrix and $\mymatrix{W}$ is an Hermitian weight matrix. The COMET estimator is defined as the minimizer of $J$:
\begin{equation}\label{eq: Definition COMET estimator}
	\myvector{\widehat \theta} = \arg \min_{\myvector{\theta}} J(\myvector{\theta}).
\end{equation}
The user-defined weight matrix $\mymatrix{W}$ may be considered as a processing parameter providing specific properties to the estimation results. In particular, $\mymatrix{W} = \mymatrix{I}$ yields an ordinary unweighted least squares estimation, while setting $\mymatrix{W} = \mymatrix{\bar R}_N^{-1}$ leads to an asymptotically efficient\footnote{if the model is well-specified}  solution \cite{ottersten1998covariance}.

As proposed in \cite{ottersten1998covariance}, \eqref{eq: Definition COMET estimator} may be solved by introducing the $M^2 \times (D + 1)$ matrix $\mymatrix{L}$ whose $k$-th column $\entry{\mymatrix{L}}_k \eqdef \vect{\mymatrix{A}_k}$, so that $\mymatrix{L}\myvector{\alpha} = \vect \mymatrix{\widehat{B}}(\myvector{\alpha})$, and the following notations:
\begin{subequations}\label{eq: Notations Phi, Psi}
\begin{align}
	\mymatrix{\Phi}(\omega_0) &\eqdef \diag \myvector{a}(\omega_0), \\
	\mymatrix{\Psi}(\omega_0) &\eqdef \mymatrix{\Phi}(\omega_0)^{\hermitian} \otimes \mymatrix{\Phi}(\omega_0), \\
	\myvector{\bar r}_N &\eqdef \vect\mymatrix{\bar R}_N.
\end{align}
\end{subequations}
Using these notations and $\myvector{\theta} = (\omega_0, \, \myvector{\alpha})$, the cost function, \eqref{eq: Definition J}, becomes:
\begin{equation}\label{eq: Cost function}
	J(\myvector{\theta}) = \frac{1}{2} \left(\myvector{\bar r}_N - \mymatrix{\Psi}(\omega_0) \mymatrix{L} \myvector{\alpha}\right)^{\hermitian} \left(\mymatrix{W}^{\intercal} \otimes \mymatrix{W}\right) \left(\myvector{\bar r}_N - \mymatrix{\Psi}(\omega_0) \mymatrix{L} \myvector{\alpha}\right).
\end{equation}

As pointed out in \cite{ottersten1998covariance}, for a fixed value of $\omega_0$, \eqref{eq: Cost function} is a quadratic form in $\myvector{\alpha}$ whose minimum is achieved at:
\begin{equation}\label{eq: Expression hat alpha}
	\myvector{\widehat \alpha}(\omega_0) \eqdef \left[(\mymatrix{\Psi}(\omega_0)\mymatrix{L})^{\hermitian}\left(\mymatrix{W}^{\intercal}\otimes \mymatrix{W}\right)\mymatrix{\Psi}(\omega_0)\mymatrix{L}\right]^{-1}(\mymatrix{\Psi}(\omega_0)\mymatrix{L})^{\hermitian}\left(\mymatrix{W}^{\intercal}\otimes \mymatrix{W}\right)\myvector{\bar r}_N.
\end{equation}
Finally, $\omega_0$ can be estimated using a simple uni-dimensional optimization of the following criterion, obtained by injecting \eqref{eq: Expression hat alpha} into \eqref{eq: Cost function}:
\begin{align}\label{eq: Optimization omega}
	\widehat \omega_0 = \arg \max& \left[(\mymatrix{\Psi}(\omega_0)\mymatrix{L})^{\hermitian}\left(\mymatrix{W}^{\intercal}\otimes \mymatrix{W}\right)\myvector{\bar r}_N\right]^{\intercal} \notag \\
	&\times \left[(\mymatrix{\Psi}(\omega_0)\mymatrix{L})^{\hermitian}\left(\mymatrix{W}^{\intercal}\otimes \mymatrix{W}\right)\mymatrix{\Psi}(\omega_0)\mymatrix{L}\right]^{-1} \notag\\
	&\times (\mymatrix{\Psi}(\omega_0)\mymatrix{L})^{\hermitian}\left(\mymatrix{W}^{\intercal}\otimes \mymatrix{W}\right)\myvector{\bar r}_N
\end{align}
The linear parameters $\myvector{\widehat \alpha}$ are then deduced by evaluating \eqref{eq: Expression hat alpha} at $\widehat \omega_0$. The parameters of interest are finally obtained as:
\begin{align}\label{eq: Expression characteristics}
		\widehat P &= \widehat \alpha_0, &
		\widehat \sigma_\omega^2 &= \widehat \alpha_1 / \widehat \alpha_0.
\end{align}

The complete estimation scheme is summarized in Algorithm~\ref{alg: Estimation algorithm}. The optimization occurs during Step 2 with the estimation of $\omega_0$. \ref{ap: Implentation aspects} details some implementation aspects that speed up this optimization.
\begin{algorithm}
	\caption{MoMET algorithm.}
	\begin{algorithmic}[1]
		\State Form the sample covariance matrix $\mymatrix{\bar{R}}_N$ from the measurements $\set{\myvector{x}(t)}$.
		\State Estimate $\widehat \omega_0$ by solving the one-dimensional optimization problem \eqref{eq: Optimization omega}.
		\State Deduce the linear parameters $\myvector{\widehat \alpha}$ from \eqref{eq: Expression hat alpha} with $\widehat\omega_0$.
		\State Deduce the source characteristics $P$ and $\sigma_\omega$ from \eqref{eq: Expression characteristics}.
	\end{algorithmic}\label{alg: Estimation algorithm}
\end{algorithm}

\section{Performance analysis}\label{sec: Performance analysis}

The covariance matrix model $\mymatrix{\widehat R}(\myvector{\theta})$ being designed by truncating the Taylor expansion of the characteristic function $\tilde p$, \eqref{eq: Parameterization characteristic function}, it is, in general, misspecified. Let $\myvector{\theta}^{\mathrm{true}}$ denote the vector of true parameters,
\begin{equation*}
	\myvector{\theta}^{\mathrm{true}} = \left( \omega_0^{\mathrm{true}}, P^{\mathrm{true}}, \myvector{\nu}^{\mathrm{true}}, \sigma_\epsilon^{\mathrm{true}}\right).
\end{equation*}
The model's misspecification has two consequences: $(i)$ there may be no parameter that reconstructs the true covariance $\mymatrix{R}$, \ie{} $\forall \myvector{\theta}$, $\mymatrix{\widehat R}(\myvector{\theta}) \ne \mymatrix{R}$ (in particular, $\mymatrix{\widehat R}(\myvector{\theta}^{\mathrm{true}}) \ne \mymatrix{R}$), and $(ii)$ even if there exists $\myvector{\theta}'$ such that $\mymatrix{\widehat R}(\myvector{\theta}') = \mymatrix{R}$, the parameters in $\myvector{\theta}'$ shall not be interpreted as the true ones, \ie{} $\myvector{\theta'} \ne \myvector{\theta}^{\mathrm{true}}$.

The choice of the expansion order, $D$, impacts the misspecification, and therefore the estimator performance. Increasing the expansion order $D$ generates finer reconstructions of the characteristic function $\tilde p$, and therefore of the covariance matrix $\mymatrix{R}$, and is expected to produce better estimates. However, the number of parameters to be estimated increases with the expansion's order. Slow estimation convergence may be experienced for higher order values. The choice of $D$ relies on a trade-off between the estimator's bias and its variance, which is discussed in the following two subsections. Subsection~\ref{ssec: Asymptotic bias} presents the convergence properties of the estimators, and Subsection~\ref{ssec: Asymptotic variance} its asymptotic covariance.

For the sake of clarity throughout this section, the number of samples $N$ is emphasized in the notation of the cost function, which is denoted by $J_N(\myvector{\theta})$. Furthermore, as the number of samples $N$ goes to infinity, $\mymatrix{\mymatrix{\bar R}_N}$ is a strongly consistent estimator of $\mymatrix{R}$. Therefore $J_N$ converge almost surely to the asymptotic cost function,
\begin{equation}\label{eq: Asymptotic cost function}
	J_\infty(\myvector{\theta}) \eqdef \frac{1}{2} \norm{\mymatrix{\widehat{R}}(\myvector{\theta}) - \mymatrix{R}}_{\mymatrix{W}}^2.
\end{equation}

Finally, it should be noted that the order $D$ is constrained by the number of sensors $M$. The parameter vector $\myvector{\theta} = \left(\omega_0,\, P,\, \myvector{\nu},\, \sigma_\epsilon^2\right)$, with $\myvector{\nu} = \left(\nu_2,\, \dots,\, \nu_D \right) \in \R^{d-1}$, consists of $d+2$ parameters to be identified. Moreover, a general linear array with $M$ sensors can observe up to $M(M-1)+1$ parameters. This number is reduced to $2M - 1$ for a ULA. Unambiguous identification requires the number of parameters to be estimated does not exceed the number of independent observations \cite{waw1996detection, ottersten1998covariance}. Therefore, the maximal possible order is given by:
\begin{equation}
	D_{\mathrm{max}} = \left\{\begin{array}{cl}
		M(M-1) -1 & \text{for a general LA,} \\
		2M - 3 & \text{for a ULA.}
	\end{array}\right.
\end{equation}

\subsection{Accumulation point and asymptotic bias}\label{ssec: Asymptotic bias}

This subsection focuses on the convergence of the MoMET estimator. To derive asymptotic properties, it is first assumed that the asymptotic cost function, $J_\infty$, has a global minimum on the ambiguity domain. That assumption is discussed at the end of that subsection. This assumption yields the following result. 

\begin{lem}
If $J_\infty$ has a global minimum, then the estimator $\myvector{\widehat\theta}$ converges almost surely to this accumulation point,
\begin{equation}\label{eq: Definition accumulation point}
	\myvector{\widehat \theta} \toas \myvector{\theta}_0 \eqdef \arg \min_{\myvector{\theta}} J_\infty(\myvector{\theta}).
\end{equation}
\end{lem}
\begin{proof}
	See \ref{ap: Proofs}.
\end{proof}
As already mentioned, because of the model's misspecification, the accumulation point, $\myvector{\theta}_0$, generally differs from the true parameter, $\myvector{\theta}^{\mathrm{true}}$, \ie{} the estimator is biased. Nonetheless, asymptotic upper bounds for the bias $\norm{\myvector{\theta}_0 - \myvector{\theta}^{\mathrm{true}}}$ can be derived as follows.

First, note that by construction of $\widehat{\tilde p}$, \eqref{eq: Parameterization characteristic function}, and $\mymatrix{\widehat R}(\myvector{\theta})$, \eqref{eq: Modeled covariance}:
\begin{equation*}
	J_\infty(\myvector{\theta}^{\mathrm{true}}) = O(\sigma_\omega^{2(D+1)}).
\end{equation*}
Furthermore, by definition of $\myvector{\theta}_0$, $0\le J_\infty(\myvector{\theta}_0) \le J_\infty(\myvector{\theta}^{\mathrm{true}})$. 
Thus, $J_\infty(\myvector{\theta}^{\mathrm{true}}) - J_\infty(\myvector{\theta}_0) = O(\sigma_\omega^{2(D+1)})$. Since $\myvector{\nabla} J_\infty(\myvector{\theta}_0) = \myvector{0}$, the Taylor expansion of $J_\infty$ around $\myvector{\theta}_0$ yields,
\begin{equation}\label{eq: DL J_infty}
	J_\infty(\myvector{\theta}^{\mathrm{true}}) - J_\infty(\myvector{\theta}_0) = (\myvector{\theta}_0 - \myvector{\theta}^{\mathrm{true}})^{\intercal} \mymatrix{H}_\infty(\myvector{\theta}_0)(\myvector{\theta}_0 - \myvector{\theta}^{\mathrm{true}}) + o(\norm{\myvector{\theta}_0 - \myvector{\theta}^{\mathrm{true}}}^2),
\end{equation}
where $\mymatrix{H}_\infty(\myvector{\theta})$ is the Hessian matrix of the cost function $J_\infty$ evaluated at $\myvector{\theta}$. Then, the following lemma allows to deduce the asymptotic behavior of the bias.
\begin{lem}\label{lem: Hessian PD}
	There exist $\sigma_0 > 0$ and $C >0$ such that $\forall \sigma_{\omega} < \sigma_0$, $\mymatrix{H}_\infty(\myvector{\theta}_0) \succeq C \mymatrix{I}$, where $\myvector{\theta}_0$ is a function of $\sigma_\omega$.
\end{lem}
\begin{proof}
	See \ref{ap: Proofs}.
\end{proof}
According to Lemma~\ref{lem: Hessian PD}, there exists a constant $C >0$ such that for $\sigma_\omega$ small enough,
\begin{equation}\label{eq: Inequality lemma}
	(\myvector{\theta}_0 - \myvector{\theta}^{\mathrm{true}})^{\intercal} \mymatrix{H}_\infty(\myvector{\theta}_0)(\myvector{\theta}_0 - \myvector{\theta}^{\mathrm{true}}) \ge C \norm{\myvector{\theta}_0 - \myvector{\theta}^{\mathrm{true}}}^2.
\end{equation}
Finally, injecting \eqref{eq: Inequality lemma} into \eqref{eq: DL J_infty} leads to $\norm{\myvector{\theta}_0 - \myvector{\theta}^{\mathrm{true}}}^2 = O(\sigma_\omega^{2(D+1)})$, and thus:
\begin{equation}\label{eq: Asymptotic bias}
	\norm{\myvector{\theta}_0 - \myvector{\theta}^{\mathrm{true}}} = O(\sigma_\omega^{D+1}).
\end{equation}

Thus, the estimation bias decreases as the source's spread decreases, and when the expansion order $D$ increases. %That observation leads to the following counter-intuitive statement: the asymptotic accuracy of the estimation of the first three moments of the density ($P$, $\omega_0$, $\sigma_\omega$) is improved by considering a higher-order model, i.e. with $D\ge3$. 
However, the significance  of this asymptotic results should be weighed against the slowdown in estimation convergence as adding parameters may require more samples to achieve similar performance, as detailed in Subsection~\ref{ssec: Asymptotic variance}.

\begin{figure*}
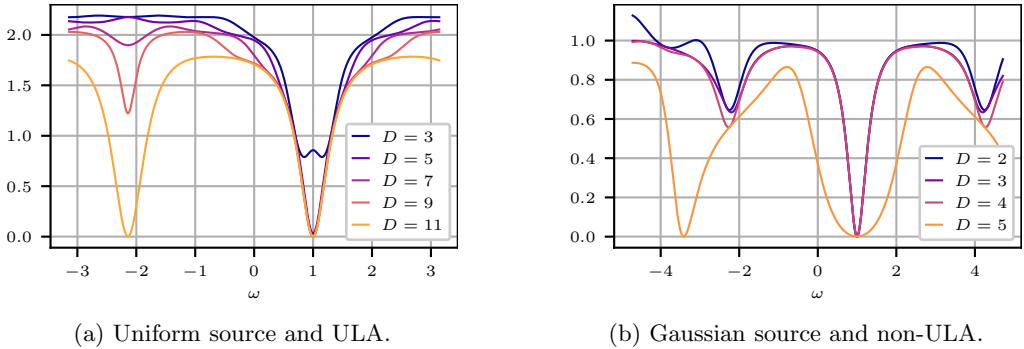

	\centering
	\null\hfill
	\subfloat[Uniform source and ULA.\label{sfig: Loss function: Uniform source ULA}]{\input{fig/loss_function_uniform_ULA.pgf}}
	\hfill
	\subfloat[Gaussian source and non-ULA.\label{sfig: Loss function: Gaussian source nonULA}]{\input{fig/loss_function_gaussian_nonULA.pgf}}
	\hfill\null
	\caption{Loss function $J_\infty(\omega, \myvector{\widehat\alpha}(\omega))$ for different orders. The same numerical values as in Fig.~\ref{fig: Comparison power densities} have been used (in particular $\omega_0 = 1$). In $(a)$ the source has an uniform distribution and the array is uniform with $7$ sensors and $u_k = k$ ($D_{\mathrm{max}}=11$). In $(b)$ the source has a Gaussian distribution and the array is non-uniform with $3$ sensors ($D_{\mathrm{max}}=5$): unevenly distributed sensor positions have been generated by adding a random zero-mean Gaussian shift with variance $\sigma^2=0.01$ to the ULA coordinates.} 
	\label{fig: Loss function}
\end{figure*}

Finally, the existence of a unique accumulation point must be discussed.
Examples of the asymptotic cost function \eqref{eq: Asymptotic cost function} are represented in Figure~\ref{fig: Loss function} for different orders $D$ and two distributions.
First, it can be observed that if the order is set to $D=D_{\mathrm{max}}$, there are several minima for which $\mymatrix{R}$ is perfectly reconstructed, \ie{} such that $J_\infty(\omega) = 0$. As this may be observed in Figure~\ref{fig: Loss function}, in the ULA case, the ambiguous minimum is located at $\omega=\omega_0 + \pi$, while in the  non-ULA the period of the ambiguity is not well defined \cite{babu2010spectral}. Such an ambiguity is recurrent with COMET method, for example it is also present in the decoupled estimation algorithm presented in \cite{besson2000decoupled, zoubir2006modified}. It can be explained as follows. The parameter $\omega_0$ represents the mean of the power density, $f(\omega)$. Since this function is periodic, shifting it by half a period also produces a zero-mean function. However, this second solution can be discarded because it yields non-physical estimates, such as a negative power.
Moreover, it can also be observed in Figure~\ref{sfig: Loss function: Uniform source ULA} that $J_\infty$ may have several minima even when $D < D_{\mathrm{max}}$, \eg{} in that figure for $D = 3$. That second observation prevents a general study. However, those minima remain in the neighborhood of $\myvector{\theta}_0$ and all the previous development still applies.

\subsection{Asymptotic covariance}\label{ssec: Asymptotic variance}

This section studies the behavior in covariance of the MoMET estimator, $\myvector{\widehat \theta}$, around its accumulation point, $\myvector{\theta}_0$, for large numbers of samples $N$. The approach used in this work follows the one described in \cite{diong2017efficiency} for the ML estimation of misspecified models.

As $\myvector{\widehat \theta}$ converges in distribution to $\myvector{\theta}_0$, a Taylor expansion of the gradient of $J_N$ around $\myvector{\theta}_0$ yields:
\begin{equation}\label{eq: DL gradient cost function}
	\myvector{0} = \myvector{\nabla} J_N(\myvector{\widehat \theta}) = \myvector{\nabla} J_N(\myvector{\theta}_0) + \mymatrix{H}_N(\myvector{\theta}_0) (\myvector{\widehat \theta} - \myvector{\theta}_0) + o\left(\norm{\myvector{\widehat \theta} - \myvector{\theta}_0}\right),
\end{equation}
where $\mymatrix{H}_N(\myvector{\theta})$ denotes the Hessian of $J_N$ evaluated at $\myvector{\theta}$.

Following the same approach as Cramér \cite[pp. 500-503]{cramer1999mathematical} and White \cite[Th. 3.2]{white1982maximum}, in the limit of large samples, \eqref{eq: DL gradient cost function} gives:
\begin{equation}
	\myvector{\widehat \theta} - \myvector{\theta}_0 \sim_N \mymatrix{H}_\infty(\myvector{\theta}_0)^{-1}\myvector{\nabla} J_N(\myvector{\theta}_0).
\end{equation}
Then, the asymptotic covariance of $\myvector{\hat \theta}$ is:
\begin{equation}\label{eq: Asymptotic expressions variance}
	\mymatrix{C}_N \eqdef \E\left[(\myvector{\widehat \theta} - \myvector{\theta}_0)(\myvector{\widehat \theta} - \myvector{\theta}_0)^{\intercal}\right] \sim_N \frac{1}{N}\mymatrix{H}_\infty(\myvector{\theta}_0)^{-1} \mymatrix{C}_0 \mymatrix{H}_\infty(\myvector{\theta}_0)^{-1},
\end{equation}
where $\mymatrix{C}_0 \eqdef \lim_{N\to\infty} N \E\left[\myvector{\nabla} J_N(\myvector{\theta}_0)\myvector{\nabla} J_N(\myvector{\theta}_0)^{\intercal}\right]$.

The details of the computation of $\mymatrix{H}_\infty(\myvector{\theta}_0)$ and $\mymatrix{C}_0$ are given in \ref{ap: Computation of the asymptotic covariance}. Their entries are:
\begin{subequations}
	\begin{align}
		\entry{\mymatrix{H}_\infty(\myvector{\theta}_0)}_{k,l} &= \trace\left\{ \mymatrix{W}\frac{\partial^2 \mymatrix{\widehat{R}}(\myvector{\theta}_0)}{\partial \theta_k \partial \theta_l} \mymatrix{W}\left(\mymatrix{\widehat{R}}(\myvector{\theta}_0)- \mymatrix{R}\right) \right\} \notag\\
		& \qquad +
		\trace\left(\mymatrix{W} \frac{\partial \mymatrix{\widehat{R}}(\myvector{\theta}_0)}{\partial \theta_k}\mymatrix{W} \frac{\partial \mymatrix{\widehat{R}}(\myvector{\theta}_0)}{\partial \theta_l}\right),\\
		\entry{\mymatrix{C}_0}_{k,l} &= \trace\left(\mymatrix{W} \frac{\partial \mymatrix{\widehat{R}}(\myvector{\theta}_0)}{\partial \theta_k} \mymatrix{W} \mymatrix{R} \mymatrix{W} \frac{\partial \mymatrix{\widehat{R}}(\myvector{\theta}_0)}{\partial \theta_l} \mymatrix{W} \mymatrix{R}\right).
	\end{align}
\end{subequations}

The expression of the asymptotic covariance, \eqref{eq: Asymptotic expressions variance}, echoes the expression of the Huber's Sandwich Bound, also known as the Misspecified Cramér-Rao Bound (MCRB), see \eg{} \cite{fortunaty2016misspecified} for an overview. Under mild conditions \cite{white1982maximum, vuong1986cramer,fortunaty2016misspecified}, the MCRB is the asymptotic covariance of the Misspecified Maximum Likelihood (MML) estimator. Despite the expected resemblance between the two covariances, which were indeed derived through a similar process, it should be noted that they are computed at two different points. The covariance matrix of $\myvector{\widehat \theta}$ \eqref{eq: Asymptotic expressions variance} is computed at accumulation $\myvector{\theta}_0$ which minimizes $J_\infty$, while the MCRB is computed at a different accumulation point which minimizes another cost function, namely the Kullback–Leibler distance between the true and the assumed model \cite{akaike1973information}. This difference implies that, unlike in the well-specified case \cite{ottersten1998covariance}, the COMET estimator is not asymptotically equivalent to the (Misspecified) ML estimator. In particular, they do not converge toward the same accumulation point since they minimize two different metrics.

\section{Numerical examples}\label{sec: Numerical example}

To illustrate the benefits of the MoMET, its performance is assessed in simulations and compared with a parametric ML estimator.

Three distributions, having a Gaussian, uniform, or exponential shape have been simulated. Unless stated otherwise, the true parameters are set to $\omega_0 = 1$~rad, $\sigma_\omega = 2\pi \times 0.05 \approx 0.31$~rad, $P = 100$~W and $\sigma_\epsilon^2 = 10$~W, and the simulated array is a ULA with $M = 7$ (the Fourier resolution of the array is $\delta_\omega = \frac{2\pi}{M} \approx 0.90$~rad). The MoMET algorithm is run with different orders, $D$, ranging from $2$ to $D_{\mathrm{max}}-1 = 10$. Two other estimators assuming a Gaussian power density are also applied for comparison: $(i)$ the ML estimator, and $(ii)$ the decoupled estimator proposed in \cite{besson2000decoupled, zoubir2006modified}. Both are well-specified and efficient for the Gaussian case, but are misspecified for the uniform and the exponential distributions. The ML estimator is expected to provide the best performance but can hardly be applied in real applications because of its high sensitivity to initialization settings. The decoupled estimator is a numerically stable alternative to the ML; that competitor has been selected because it has a similar complexity as the MoMET.

\begin{figure*}
	\centering
	\input{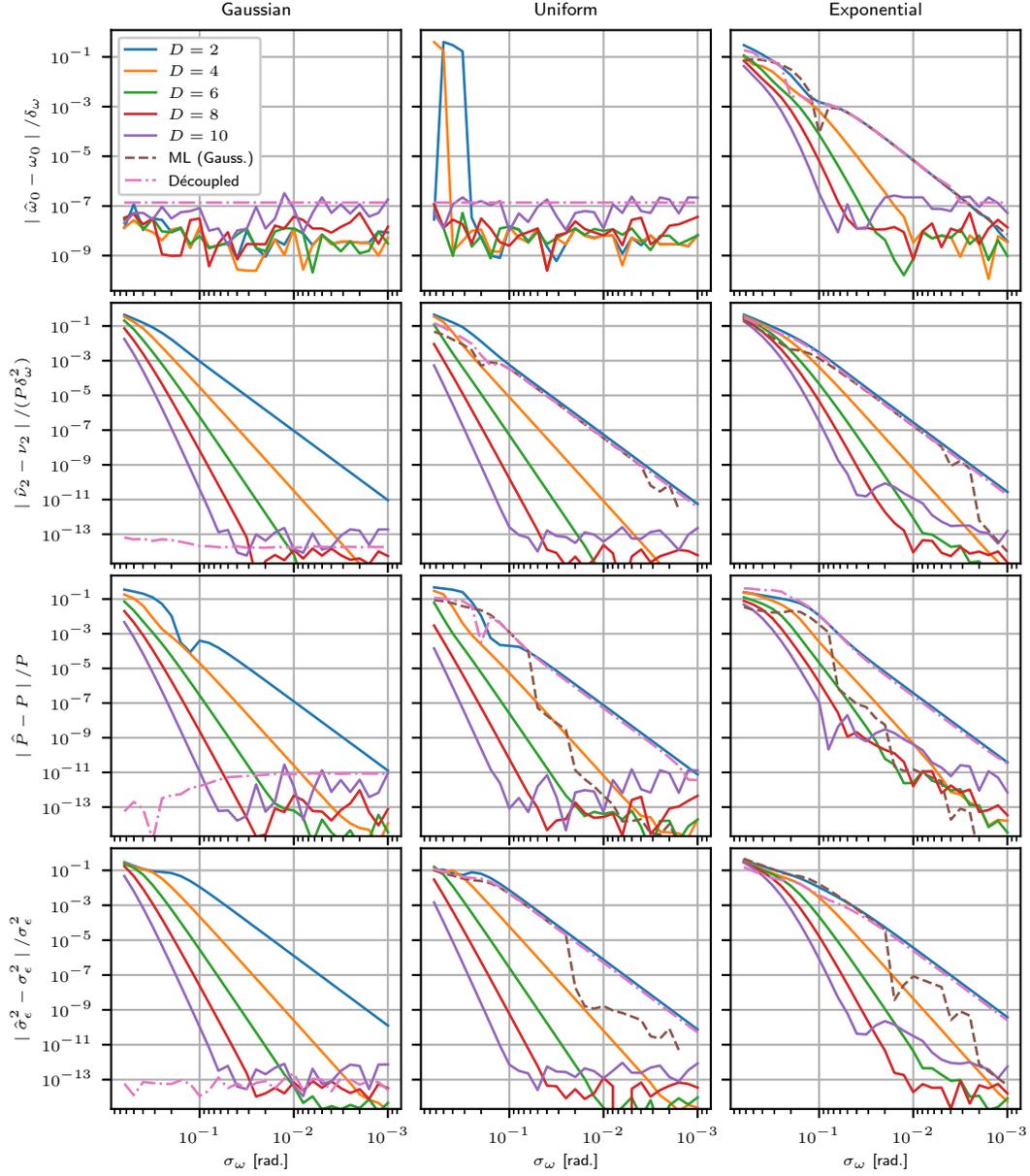}
	\caption{Normalized asymptotic biases as a function of the spread $\sigma_\omega$ for three shaping distributions. The accumulation points $\myvector{\theta}_0$ were obtained by minimizing $J_\infty(\myvector{\theta})$.}
	\label{fig: Asymptotic biases}
\end{figure*}
Figure~\ref{fig: Asymptotic biases} presents the evolution of the asymptotic bias for different orders as a function of the source's spread $\sigma_\omega$.
For the MoMET, the biases reduce as the spread $\sigma_\omega$ reduces, and more precisely they are $O(\sigma_\omega^{D+1})$, as stated in Section~\ref{ssec: Asymptotic bias}, \eqref{eq: Asymptotic bias}. As expected, larger orders result in better asymptotic performance. For the Gaussian distribution and the uniform distribution, the biases on $\omega_0$ reach the system precision ($\sim 10^{-8}$ for this parameter) which suggests that these estimators are unbiased. This may be due to the symmetry of these distributions.

\begin{figure*}
	\centering
	\input{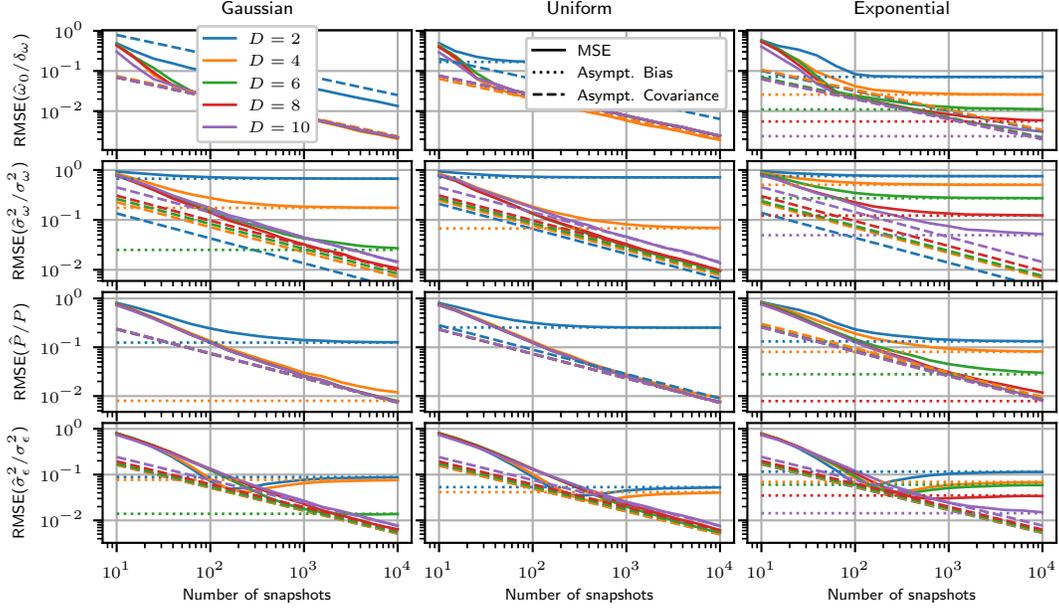}
	\caption{Normalized RMSE computed over $1\ 000$ Monte Carlo runs for different orders. The dashed lines represent the asymptotic covariances, the dotted lines the asymptotic biases.}
	\label{fig: RMSE}
\end{figure*}

\begin{figure*}
	\centering
	\input{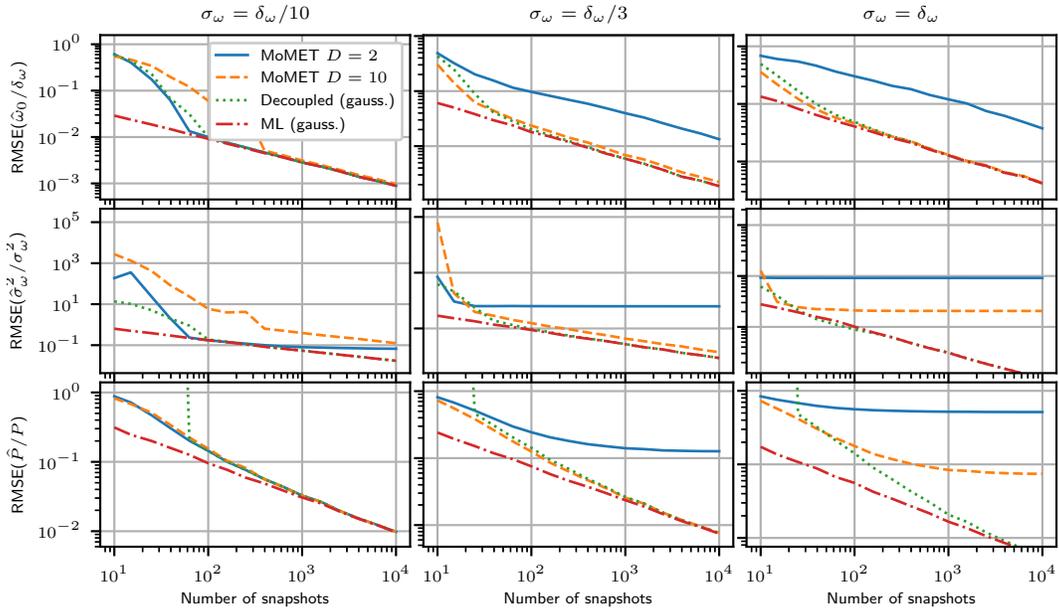}
	\caption{Normalized RMSE computed over $1\ 000$ Monte Carlo runs for a source having a Gaussian density. The competitors are well-specified.}
	\label{fig: RMSE Gaussian}
\end{figure*}

\begin{figure*}
	\centering
	\input{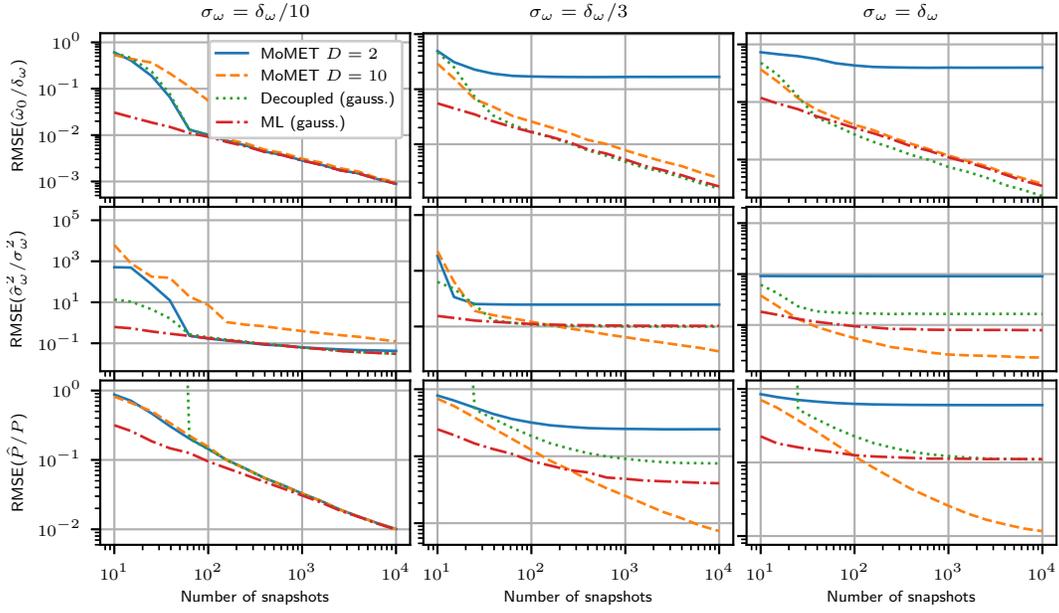}
	\caption{Normalized RMSE computed over $1\ 000$ Monte Carlo runs for a source having a uniform density. The competitors are misspecified.}
	\label{fig: RMSE Uniform}
\end{figure*}

\begin{figure*}
	\centering
	\input{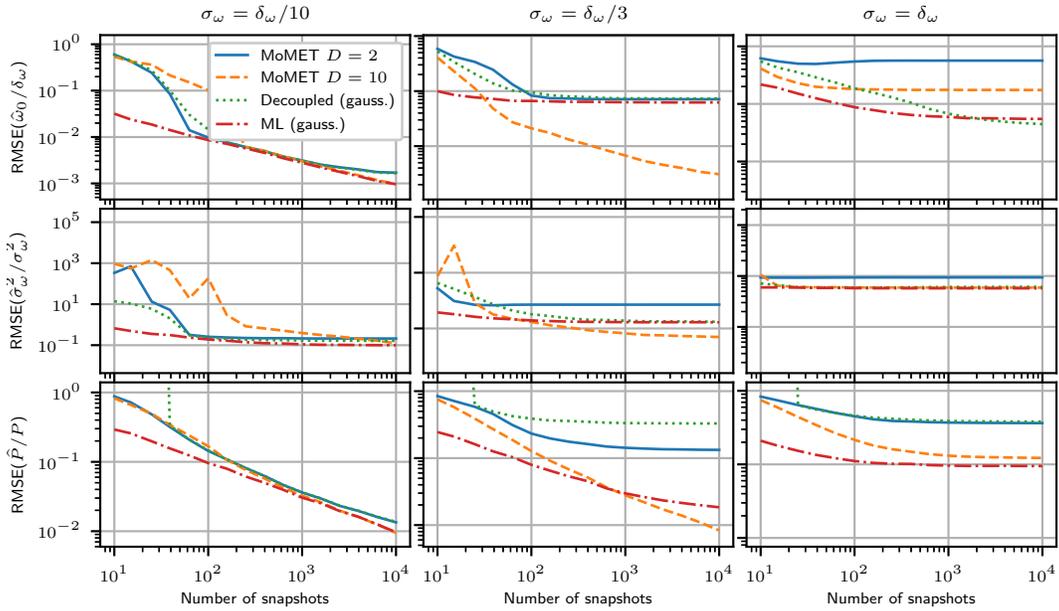}
	\caption{Normalized RMSE computed over $1\ 000$ Monte Carlo runs for a source having an exponential density. The competitors are misspecified.}
	\label{fig: RMSE Exponential}
\end{figure*}

To assess the practical performance of the estimators, the Root Mean Square Error (RMSE) has been computed over $1\ 000$ Monte Carlo runs for different numbers of snapshots $N$.
Figure~\ref{fig: RMSE} compares the performance of the MoMET applied with different orders, $D$. That figure also represents the asymptotic covariance and the asymptotic biases. It can be observed that larger order, $D$, present larger covariances. However, as they also present smaller bias, they reach eventually, for large number of samples, better RMSE performance. This illustrates the well-known trade-off in estimation between bias and variance \cite{hero1996exploring}.
Figures~\ref{fig: RMSE Gaussian}-\ref{fig: RMSE Exponential} compare the performance of the MoMET estimator with the competitors on the estimation of the three parameters of interest. Each figure presents the result for one of the three considered distributions and three different source's spreads. For clarity, only the MoMET with minimal order, $D=2$, and maximal order, $D=D_{\mathrm{max}}-1 = 10$, are shown in the figures. When the source's spread is small compared with the Fourier resolution (\eg{} $\sigma_\omega = \delta_\omega/10$), small order MoMET performs as well as the decoupled estimation. In this regime, the actual shape of the distribution does not impact the estimation performance. For larger spreads, applying MoMET with higher orders produce better results, indicating that a second-order approximation is no longer sufficient. The choice of the order should therefore be influenced by the source's spread. Of course, the competitors perform better when their distribution assumption is correct, \ie{} for the Gaussian case, Fig.~\ref{fig: RMSE Gaussian}. However, for the two misspecified cases, Figs.~\ref{fig: RMSE Uniform} and \ref{fig: RMSE Exponential}, MoMET reaches similar or better performance than the competitors, especially for the estimation of the spread.
%The two competitors are unbiased for the Gaussian distribution, but are biased for the two misspecified cases as expected. Generally, the MoMET applied with higher orders obtains better bias than its competitors.

\section{Conclusion}\label{sec: Conclusion}

This paper has presented a new estimation scheme, called MoMET, for the characterization of a diffuse source. In order to avoid incorrect models, the MoMET scheme does not assume any particular shape for the source's distribution. Instead, the unknown distribution is characterized by its mean DOA and its first central moments. The estimation is then realized by applying the COMET procedure. A performance analysis has derived the asymptotic biases and asymptotic covariance of the MoMET estimator. It has been highlighted that estimating more moments results in a global gain of performance, at the cost of a slower convergence rate. This performance gain has been validated through simulations. The proposed algorithm achieved similar or better performance than a misspecified ML estimator. This work demonstrates how sometimes, using an unrealistic and misspecified but generic model can be better than having an incorrect realistic model.

The proposed model is easy to implement and to optimize: the cost function is a quadratic and all parameters, but one, enter linearly in the expression of the solution. Therefore, the proposed algorithm has a low numerical complexity. Moreover, it can be applied with both uniform or non-uniform linear arrays. All those points make it particularity suitable for large scale imagery applications such as forest SAR tomography.

Several avenues are possible for the future. First, the extension of the MoMET to several sources shall be studied. On a theoretical level, questions remain open and will require further investigation: for example, the absence of bias in the case of symmetric distributions, or the choice of the hyper-parameter. Finally, the upcoming arrival of data from the Biomass mission \cite{quegan2019european} will enable the proposed algorithms to be applied and validated on real data.

\appendix

\section{Implementation aspects}\label{ap: Implentation aspects}

This appendix discusses numerical aspects of the estimation algorithm, and specifically, the estimation of $\omega_0$ in \eqref{eq: Definition COMET estimator}, which represent the most demanding part of the optimization algorithm. In practice, the algorithm is applied a large number of times with different data. The following paragraphs propose to speed up the maximization by reformulating and precalculating some terms.

The optimization problem described in \eqref{eq: Optimization omega} may be reformulated as  $\widehat \omega_0 = \arg \max \myvector{y}(\omega)^{\hermitian} \mymatrix{Y}(\omega)^{-1}$ with: 
\begin{subequations}	
	\begin{align}
		\mymatrix{Y}(\omega) &\eqdef (\mymatrix{\Psi}(\omega)\mymatrix{L})^{\hermitian}\left(\mymatrix{W}^{\intercal}\otimes \mymatrix{W}\right)\mymatrix{\Psi}(\omega)\mymatrix{L}, \\
		\myvector{y}(\omega) &\eqdef \mymatrix{L}^{\hermitian} \mymatrix{\Psi}(\omega)^{\hermitian}\left(\mymatrix{W}^{\intercal}\otimes \mymatrix{W}\right)\myvector{\bar r}_N,
	\end{align}
\end{subequations}
where $\mymatrix{\Psi}(\omega)$ is defined in \eqref{eq: Notations Phi, Psi}, $\mymatrix{L}$ is the matrix of linear parameters introduced in Section~\ref{ssec: Estimation algorithm}, and $\mymatrix{W}$ is an Hermitian weighting matrix.

\paragraph{On the computation of $\myvector{y}(\omega)$} 
First, note that for any matrix $\mymatrix{Z}$:
\begin{equation*}
	\left[\mymatrix{L}^{\hermitian} \vect(\mymatrix{Z})\right]_k = \trace{\mymatrix{A}_k \mymatrix{Z}},
\end{equation*}
with $\mymatrix{A}_k$ explicited in \eqref{eq: Matrices linear parameters}.
%Therefore, if $\mymatrix{Y}$ is Hermitian, $\mymatrix{L}^{\hermitian} \mymatrix{Y} \in \R^m$.
In particular as:
\begin{align*}
	\myvector{y}(\omega) &= \mymatrix{L}^{\hermitian} \mymatrix{\Psi}(\omega)^{\hermitian} \left(\mymatrix{W}^{\intercal} \otimes \mymatrix{W}\right) \myvector{\bar r}_N, \\
	&= \mymatrix{L}^{\hermitian} \vect\left\{ \mymatrix{\Phi}(\omega)^{\hermitian} \mymatrix{W}\mymatrix{\bar R}_N\mymatrix{W} \mymatrix{\Phi}(\omega)\right\},
\end{align*}
the $k$-th entry of $\myvector{y}(\omega)$ is:
\begin{align*}
	\left[\myvector{y}(\omega)\right]_k &= \trace \mymatrix{A}_k \mymatrix{\Phi}(\omega)^{\hermitian} \mymatrix{W}\mymatrix{\bar R}_N\mymatrix{W} \mymatrix{\Phi}(\omega), \\
	&= \myvector{a}(\omega)^{\hermitian} \left( \mymatrix{W}\mymatrix{\bar R}_N\mymatrix{W} \odot \mymatrix{A}_k^{\intercal} \right) \myvector{a}(\omega).
\end{align*}
The matrices $\left( \mymatrix{W}\mymatrix{\bar R}_N\mymatrix{W} \odot \mymatrix{A}_k^{\intercal} \right)$ are independent from $\omega$, and can be precalculated to speed up the optimization of $\widehat \omega_0$.

Furthermore, in the special case of a ULA, the computation can be further simplified, as:
\begin{align*}
	\left[\myvector{y}(\omega)\right]_k &=  \myvector{a}(\omega)^{\hermitian} \left( \mymatrix{W}_R \odot \mymatrix{A}_k^{\intercal} \right) \myvector{a}(\omega), \\
	&= \sum_{m, n} \entry{\mymatrix{A}_k}_{n, m} \entry{\mymatrix{W}_R}_{m,n} e^{j (n-m)\omega},
\end{align*}
where $\mymatrix{W}_R \eqdef \mymatrix{W}\mymatrix{\bar R}_N\mymatrix{W}$.
Then, by grouping the terms along the diagonals, and noting that the entries of the $\set{\mymatrix{A}_k}$ are constant along the diagonals, one gets:
\begin{align*}
	\left[\myvector{y}(\omega)\right]_k &= \entry{\mymatrix{A}_k}_{1,1} \trace \mymatrix{W}_R \\
	&\qquad + 2 \Real \left\{\sum_{l=1}^{M-1} \entry{\mymatrix{A}_k}_{l+1, 1} \left(\sum_{m = 1}^{M-l}\entry{\mymatrix{W}_R}_{m, m+l}\right) e^{j l\omega}\right\}.
\end{align*}
Thus, there exists a precomputable matrix $\mymatrix{B}$ such that:
\begin{equation*}
	\myvector{y}(\omega) = \Real\left\{\mymatrix{B} \myvector{a}(\omega)\right\}.
\end{equation*}
Let $\myvector{w} = \begin{pmatrix} w_0 & \cdots & w_{M-1}\end{pmatrix}^{\intercal}\in \C^{M}$ be the vector containing the sum of the diagonals of $\mymatrix{W}_R$. The matrix $\mymatrix{B}$ is defined as:
\begin{equation*}
	\entry{\mymatrix{B}}_{k,l} = (1 + \delta(l,0)) \entry{\mymatrix{A}_k}_{l,1} w_l.
\end{equation*}

\paragraph{On the computation of $\myvector{Y}(\omega)$}

Recall that by definition:
\begin{align*}
	\entry{\mymatrix{Y}(\omega)}_{k,l} &= \vect{(\mymatrix{A}_k)}^{\hermitian} \mymatrix{\Psi}(\omega)^{\hermitian} \left(\mymatrix{W}^{\intercal} \otimes \mymatrix{W} \right) \mymatrix{\Psi}(\omega) \vect{(\mymatrix{A}_l)}, \\
	&= \trace\left\{\mymatrix{A}_k \mymatrix{\Phi}(\omega)^{\hermitian}\mymatrix{W}\mymatrix{\Phi}(\omega) \mymatrix{A}_l \mymatrix{\Phi}(\omega)^{\hermitian} \mymatrix{W} \mymatrix{\Phi}(\omega)\right\}.
\end{align*}
Although $\mymatrix{Y}(\omega)$ cannot be simplified in the general case, this is possible in common situations. In the special case of unweighted estimation, \ie{} $\mymatrix{W} = \mymatrix{I}$, $\mymatrix{Y}(\omega)$ is independent from $\omega$ and $\mymatrix{\bar{R}}$:
\begin{equation*}
	\entry{\mymatrix{Y}}_{k,l} = \trace{\mymatrix{A}_k\mymatrix{A}_l}.
\end{equation*}

Furthermore, in the case of a ULA it can be shown that there exists a tensor $\mymatrix{C}= ((c_{k,l,m}))$ such that, for all $k$, $l$:
\begin{equation*}
	\left[\mymatrix{Y}(\omega)\right]_{k,l}  = \sum_{m = -2M + 2}^{2M-2} c_{k,l,m} e^{j m \omega}.
\end{equation*}
The entries $\set{c_{k,l,m}}$ depend only on $\mymatrix{W}$ and the $\set{\mymatrix{A}_k}$ and can also be precomputed.

\section{Proof of the lemmas}\label{ap: Proofs} 

\begin{lem*}
	If $J_\infty$ has a global minimum, then the estimator $\myvector{\widehat\theta}$ converges almost surely to this accumulation point,
	\begin{equation*}
		\myvector{\widehat \theta} \toas \myvector{\theta}_0 \eqdef \arg \min_{\myvector{\theta}} J_\infty(\myvector{\theta}).
	\end{equation*}
\end{lem*}
\begin{proof}
	Let us prove that $\lim_{N\to \infty} \myvector{\widehat \theta} = \myvector{\theta}_0$ with probability $1$. For the sake of clarity, let us emphasize on the number of samples in the definition of $\myvector{\widehat \theta}$ : $\myvector{\widehat \theta}_N = \arg\min_\theta J_N(\myvector{\theta})$.
	
	Let $\varepsilon > 0$, consider an outcome and let us prove that there exists $N_0$ such that $\forall N\ge N_0$, $\norm{\myvector{\theta_0} - \myvector{\widehat \theta}_N} \le \varepsilon$. First, note that the search space is a compact set; indeed $\myvector{\widehat \theta}$ is parameterized only by $\widehat \omega_0$ which belongs to the ambiguity domain, \eg{} $\widehat \omega_0 \in [-\pi, \pi]$ for a ULA.
	
	Since $J_\infty$ has a unique minimum, $\theta_0$, and $\mymatrix{\widehat R}$ is a continuous function of $\myvector{\theta}$, there exists $\eta > 0$ such that, $\forall \myvector{\theta}$:
	\begin{equation*}
		\abs{J_\infty(\myvector{\theta}) - J_\infty(\myvector{\theta}_0)} \le \eta \implies \norm{\myvector{\theta} - \myvector{\theta_0}} \le \varepsilon.
	\end{equation*}
	
	Moreover, note that for all $\myvector{\theta}$:
	\begin{equation}\label{eq: Link J_N, J_infty}
		J_N(\myvector{\theta}) = J_\infty(\myvector{\theta}) + \frac{1}{2}\norm{\mymatrix{R} - \mymatrix{\bar R}_N}^2_W + \langle \mymatrix{\bar R}_N - \mymatrix{R} \,, \mymatrix{R} - \mymatrix{\widehat R}(\myvector{\theta})\rangle_W,
	\end{equation}
	where $\langle \cdot\,, \cdot \rangle_W$ denotes the scalar product induced by $\mymatrix{W}$. As $\mymatrix{\widehat R}(\myvector{\theta})$ is continuous, $\myvector{\theta}$ belongs to a compact set, $\mymatrix{R} - \mymatrix{\widehat R}(\myvector{\theta})$ is bounded. Furthermore, as $\mymatrix{\bar R}_N$ converges almost surely toward $\mymatrix{R}$ according to the Central-Limit Theorem, \eqref{eq: Link J_N, J_infty} yields that $J_N$ converges almost surely toward $J_\infty$. Therefore, there exists $N_0$ such that $\forall \theta$, $\forall N \ge N_0$, $\abs{J_N(\myvector{\theta}) - J_\infty(\myvector{\theta})} \le \eta/2$. Applying twice that inequality at $\myvector{\widehat \theta}_N$ and $\myvector{\theta}_0$ gives that for all $N \ge N_0$:
	\begin{equation*}
		J_\infty(\myvector{\theta_0}) \le J_\infty(\myvector{\widehat\theta}_N) \le J_N(\myvector{\widehat\theta}_N) + \frac{\eta}{2} \le J_N(\myvector{\widehat\theta}_N) + \frac{\eta}{2} \le J_N(\myvector{\theta}_0) + \eta.
	\end{equation*}
	Therefore, $\abs{ J_\infty(\myvector{\widehat\theta}_N) -J_\infty(\myvector{\theta_0})} \le \eta$, and thus $\norm{\myvector{\widehat\theta}_N - \myvector{\theta_0}} \le \varepsilon$ as claimed.
	
\end{proof}

\begin{lem*}
	For almost every linear array, and in particular for a ULA, there exist $\sigma_0 > 0$ and $\varepsilon >0$ such that $\forall \sigma_{\omega} < \sigma_0$, $\mymatrix{H}_\infty(\myvector{\theta}_0) \succeq \varepsilon \mymatrix{I}$, where $\myvector{\theta}_0$ is implicitly seen as a function of $\sigma_\omega$.
\end{lem*}
\begin{proof}
	Consider the limit case where $\sigma_\omega = 0$, \ie{} the shaping distribution is the Dirac delta function. In that case, the model is not misspecified since $\myvector{\theta}_0 = \left(\omega_0^{\mathrm{true}},\, P^{\mathrm{true}},\, \myvector{0},\, \sigma_\epsilon^{2\mathrm{true}}\  \right) = \myvector{\theta}^{\mathrm{true}}$, and $\mymatrix{R} = P \myvector{a}(\omega_0)\myvector{a}(\omega_0)^{\hermitian} + \sigma_\epsilon^2\mymatrix{I} = \mymatrix{\widehat R}(\myvector{\theta}_0)$.
	
	In that well-specified case, the entries of the Hessian matrices are (the computation is detailed in \ref{ap: Computation of the asymptotic covariance}):
	\begin{equation}\label{eq: Hermitian well-specified}
		\entry{\mymatrix{H}_\infty(\myvector{\theta}_0)}_{k,l} =
		\trace\left\{
		\mymatrix{W}\frac{\partial \mymatrix{\hat R}(\myvector{\theta}_0)}{\partial \theta_k}\mymatrix{W}\frac{\partial \mymatrix{\hat R}(\myvector{\theta}_0)}{\partial \theta_l}
		\right\}.
	\end{equation}
	
	The derivatives of $\mymatrix{\widehat R}(\theta)$ are:
	\begin{equation*}
		\frac{\partial \mymatrix{\hat R}(\myvector{\theta}_0)}{\partial \theta_k} = \myvector{a}(\omega_0)\myvector{a}(\omega_0)^{\hermitian} \odot \mymatrix{A}_k,
	\end{equation*}
	where the matrices $\mymatrix{A}_k$ are given in \eqref{eq: Matrices linear parameters} for the linear parameters, and $\entry{\mymatrix{A}_0}_{k,l} = j(u_k-u_l)$ for the parameter $\omega_0$.
	Therefore, \eqref{eq: Hermitian well-specified} can be reexpressed as:
	\begin{equation*}
		\entry{\mymatrix{H}_\infty(\myvector{\theta}_0)}_{k,l} =
		\trace\left\{
		\mymatrix{\Phi}(\omega_0)^{\hermitian}\mymatrix{W}\mymatrix{\Phi}(\omega_0)\mymatrix{A}_k\mymatrix{\Phi}(\omega_0)^{\hermitian}\mymatrix{W}\mymatrix{\Phi}(\omega_0)\mymatrix{A}_l
		\right\},
	\end{equation*}
	which is the inner product induced by $\mymatrix{\Phi}(\omega_0)^{\hermitian}\mymatrix{W}\mymatrix{\Phi}(\omega_0)$. So, the Hermitian matrix is the Gram matrix associated with the family of matrices $\mymatrix{A}_k$. This family is linearly independent (it is related with the Vandermonde matrix), so $\mymatrix{H}_\infty(\myvector{\theta}_0)$ is positive definite. The result is finally obtained by continuity as $\sigma_\omega$ tends to $0$.
	
\end{proof}

\section{Computation of the asymptotic covariance}\label{ap: Computation of the asymptotic covariance}

Recall the expression of the asymptotic cost function, \eqref{eq: Asymptotic cost function}:
\begin{equation*}J_\infty(\myvector{\theta}) = \frac{1}{2} \trace\left\{
	\mymatrix{W}(\mymatrix{\hat R}(\myvector{\theta}) - \mymatrix{R})\mymatrix{W}(\mymatrix{\hat R}(\myvector{\theta}) - \mymatrix{R})
	\right\}.
\end{equation*}
Then, the gradient and the Hessian of $J_\infty$ are:
\begin{align*}
	\frac{\partial J_\infty (\myvector{\theta})}{\partial \theta_k} &= \trace\left\{
	\mymatrix{W}\frac{\partial \mymatrix{\hat R}(\myvector{\theta})}{\partial \theta_k}\mymatrix{W}(\mymatrix{\hat R}(\myvector{\theta}) - \mymatrix{R})
	\right\}, \\
	\entry{\mymatrix{H}_\infty(\myvector{\theta})}_{k,l} &=
	\trace\left\{
	\mymatrix{W}\frac{\partial^2 \mymatrix{\hat R}(\myvector{\theta})}{\partial \theta_k \partial \theta_l}\mymatrix{W}(\mymatrix{\hat R}(\myvector{\theta}) - \mymatrix{R})
	\right\} \\
	&\qquad + \trace\left\{
	\mymatrix{W}\frac{\partial \mymatrix{\hat R}(\myvector{\theta})}{\partial \theta_k}\mymatrix{W}\frac{\partial \mymatrix{\hat R}(\myvector{\theta})}{\partial \theta_l}
	\right\}.
\end{align*}

Similarly, the gradient of $J_N$ is:
\begin{align*}
	\frac{\partial J_N (\myvector{\theta})}{\partial \theta_k} &= \trace\left\{
	\mymatrix{W}\frac{\partial \mymatrix{\hat R}(\myvector{\theta})}{\partial \theta_k}\mymatrix{W}(\mymatrix{\hat R}(\myvector{\theta}) - \mymatrix{\bar R}_N)
	\right\}, \\
	 &=
	\trace\left\{
	\mymatrix{W}\frac{\partial \mymatrix{\hat R}(\myvector{\theta})}{\partial \theta_k}\mymatrix{W}\mymatrix{\hat R}(\myvector{\theta})
	\right\} \\
	&\qquad -
	\frac{1}{N} \sum_t
	\myvector{x}(t)^{\hermitian} \mymatrix{W}\frac{\partial \mymatrix{\hat R}(\myvector{\theta})}{\partial \theta_k}\mymatrix{W}\myvector{x}(t)
\end{align*}
Therefore,
\begin{align*}
	&\frac{\partial J_N (\myvector{\theta})}{\partial \theta_k} \frac{\partial J_N (\myvector{\theta})}{\partial \theta_l} = \trace\left\{
	\mymatrix{W}\frac{\partial \mymatrix{\hat R}(\myvector{\theta})}{\partial \theta_k}\mymatrix{W}(\mymatrix{\hat R}(\myvector{\theta}) - \mymatrix{\bar R}_N)
	\right\} \\
	&  \qquad \times\trace\left\{
	\mymatrix{W}\frac{\partial \mymatrix{\hat R}(\myvector{\theta})}{\partial \theta_l}\mymatrix{W}\mymatrix{\hat R}(\myvector{\theta})
	\right\} \\
	&\qquad - \trace\left\{
	\mymatrix{W}\frac{\partial \mymatrix{\hat R}(\myvector{\theta})}{\partial \theta_k}\mymatrix{W}\mymatrix{\hat R}(\myvector{\theta})
	\right\}
	 \trace\left\{
	\mymatrix{W}\frac{\partial \mymatrix{\hat R}(\myvector{\theta})}{\partial \theta_l}\mymatrix{W}\mymatrix{\bar R}_N
	\right\} \\
	&\qquad + \frac{1}{N^2} \sum_{t,t'}
	\myvector{x}(t)^{\hermitian} \mymatrix{W}\frac{\partial \mymatrix{\hat R}(\myvector{\theta})}{\partial \theta_k}\mymatrix{W}\myvector{x}(t)
	\myvector{x}(t')^{\hermitian} \mymatrix{W}\frac{\partial \mymatrix{\hat R}(\myvector{\theta})}{\partial \theta_l}\mymatrix{W}\myvector{x}(t').
\end{align*}
Using the fact that the $\myvector{x}(t)$ are i.i.d. and the following property,
\begin{equation*}
	\E\left\{{\myvector{x}(t)^{\hermitian}\myvector{A}x(t)\myvector{x}(t)^{\hermitian}\mymatrix{B}\myvector{x}(t)}\right\} = \trace\left\{\mymatrix{R}\mymatrix{A}\right\}\trace\left\{\mymatrix{R}\mymatrix{B}\right\} + \trace\left\{\mymatrix{R}\mymatrix{A}\mymatrix{R}\mymatrix{B}\right\},
\end{equation*}
yields
\begin{align*}
	&\E\left\{\frac{\partial J_N (\myvector{\theta})}{\partial \theta_k} \frac{\partial J_N (\myvector{\theta})}{\partial \theta_l}\right\} = 
	\frac{\partial J_\infty (\myvector{\theta})}{\partial \theta_k}
	\frac{\partial J_\infty (\myvector{\theta})}{\partial \theta_l}\\
	&\qquad + \frac{1}{N} \trace\left(\mymatrix{W} \frac{\partial \mymatrix{\widehat{R}}(\myvector{\theta})}{\partial \theta_k} \mymatrix{W} \mymatrix{R} \mymatrix{W} \frac{\partial \mymatrix{\widehat{R}}(\myvector{\theta})}{\partial \theta_l} \mymatrix{W} \mymatrix{R}\right).
\end{align*}
Finally, since by definition of $\myvector{\theta}_0$, $\myvector{\nabla} J_\infty(\myvector{\theta}_0) = \myvector{0}$, the first term cancels out and \eqref{eq: Asymptotic expressions variance} is obtained.

\bibliographystyle{elsarticle-num} 
\bibliography{bibliographie}

\end{document}